# Proton Induced Reactions on $^{114}$Sn and $^{120}$Sn Targets at Energies up to 18 MeV

G.H. Hovhannisyan[a,*], T.M. Bakhshiyan[b], G.V. Martirosyan[a], R.K. Dallakyan[c], A.R. Balabekyan[a]

[a] *Yerevan State University, 0025, Armenia*
[b] *Institute of Applied Problems of Physics, Yerevan 0014, Armenia*
[c] *A. Alikhanyan National Science Laboratory, Yerevan 0036, Armenia*

ABSTRACT

We measure cross sections of proton-induced reactions on tin up to energies of 18 MeV using the stacked-foil activation technique, and report first experimental values for $^{114}$Sn(p,α)$^{111}$In, $^{120}$Sn(p,α)$^{117g}$In, and $^{114}$Sn(p,x)$^{113}$Sn reactions. Measured cross sections have been compared to existing experimental values and numerical calculations based on Talys1.95. Our data are in good agreement with all previous experiments, and also with Talys1.95 results except for $^{120}$Sn(p,α)$^{117m,g}$In reactions where the numerical calculations are shifted to higher energies. We also measure isomeric cross section ratios for $^{117m,g}$In and $^{120m,g}$Sb pairs as functions of the incident proton energy.

## 1. Introduction

Excitation functions of proton-induced nuclear reactions are of interest in a variety of fields, ranging from astrophysical applications to medical radioisotope production [1]. The knowledge of the excitation functions allows one to approximately determine the production yields of radionuclides and to estimate the impurity fractions. There are microscopic and phenomenological models for calculating the cross-sections of nuclear reactions for energy ranges of interest, but full microscopic understanding of nuclear interactions remains an open problem.

Obtaining experimental data on nuclear reaction cross-sections is crucial for verifying the predictions of existing models and refining their parameters.

The emission of one or more nucleons in the interactions of light projectiles with nuclei is generally well described by the existing models. The emission of complex particles, such as alpha particles, is usually more difficult to predict. When analyzing (p,α) reaction cross sections, three components are usually separated: direct, pre-equilibrium, and compound, and each of them can be analyzed by the appropriate theory. The analysis of the compound-nucleus component may be made with statistical theories that require only the nuclear level densities and the transmission coefficients of the incoming and outgoing particles. Direct and pre-equilibrium processes allow the study of more fundamental questions, since their characteristics depend on the reaction mechanism and the structure of the nucleus. Triton capture and knock-out of α-particles are considered to be the main mechanisms of the direct processes. The α-particle knock-out mechanism could provide a powerful probe of the α-particle cluster structure of nuclei [2]. There is evidence indicating the difference in the contribution of direct processes depending on the nucleon composition of the target-nucleus; in particular, it is greater for neutron-rich nuclei [3].

In this article we investigate proton-induced nuclear reactions on $^{114}$Sn and $^{120}$Sn targets for energies up to 18 MeV. We measure cross sections of $^{114}$Sn(p,α)$^{111}$In and $^{120}$Sn(p,α)$^{117m,g}$In, $^{114}$Sn(p,2n)$^{113}$Sb, $^{120}$Sn(p,n)$^{120m,g}$Sb, and $^{114}$Sn(p,x)$^{113}$Sn reactions. Tin has a magic proton number (Z=50) and 10 stable isotopes with different deformation by neutrons. The knowledge of activation cross section of tin isotopes can be essential for understanding the processes governing the reactions and clarifying model parameters. However, the number of experimental works on enriched tin targets is not large. There are measurements on natural composition tin targets in the energy range covering the excitation functions [4-6]. Most measurements of proton nuclear reaction on enriched tin samples in the peak region are for energies up to 9.1 MeV and do not cover the entire excitation function range [7-17,19-23] with only a few with measurements at higher energies: for $^{124}$Sn(p,n)$^{124}$Sb reaction there are measurements for energies up to 17 MeV [24,25], for $^{112}$Sn(p,x) – up to 23.6 MeV [26], for $^{119}$Sn(p,n), $^{119}$Sn(p,2n) – up to 15.2 MeV [27], and for $^{120}$Sn(p,α) – up to 22 MeV [18].

## 2. Experimental details

We irradiated a stack of enriched $^{114}$Sn and $^{120}$Sn foils using 18 MeV proton beam provided by compact medical cyclotron IBA Cyclone18/18. The stack was composed of 6 blocks of $^{nat}$Cu-$^{114}$Sn-$^{nat}$Cu-$^{120}$Sn layers where tin foils were 20 to 40 μm thick and copper foils were 20 μm thick. The irradiation was 5 min long with a collimated 1 μA proton beam of the same diameter as the target (1.2 cm).

After the irradiation the foils in the stack were detached, and the γ-spectra of each target were measured with a high-purity germanium (HPGe) detector GEM15P4-70. The energy resolution of the HPGe detector was 1.66 keV FWHM at 1332.5 keV peak of $^{60}$Co, and 0.618 keV FWHM at 122 keV peak of $^{57}$Co. The efficiency of the detector was estimated using standard γ-sources $^{133}$Ba, $^{137}$Cs, $^{60}$Co, $^{22}$Na with known activities (supplied by Spectrum Techniques, USA), covering the whole energy range of the studied γ-rays. The measurements were performed periodically during several days, with the first measurement done 40 minutes after the end of irradiation (to allow short-lived isotope cross-section measurements). Residual nuclei were identified by their half-lives and the energies of characteristic gamma lines. A typical γ-spectrum of the irradiated $^{120}$Sn target is plotted in Fig.1.

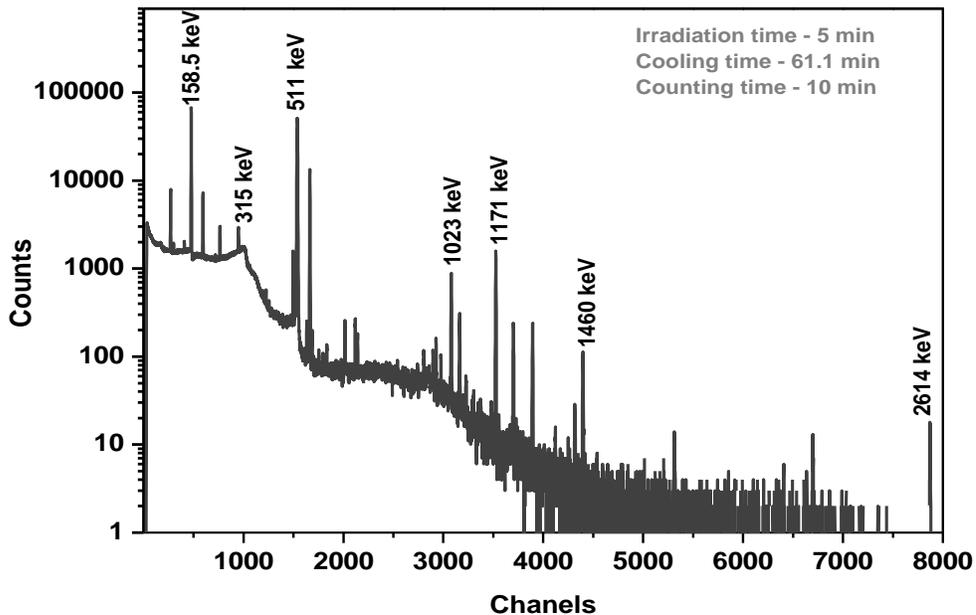

**Fig. 1.** γ–ray spectrum of the irradiated $^{120}$Sn.

The isotopic compositions of the targets are given in Table 1. Reaction thresholds, Q-values, half-lives, and branching intensities of the identified radioactive residuals formed in the $^{114}$Sn, $^{120}$Sn targets are given in Table 2. Q-values and the threshold energies of the reactions were calculated using the Q-value calculator [28], and the decay data were taken from [29-31].

**Table 1**
Isotopic composition of the targets.

| Target | Isotopic composition (%) | | | | | | | | | |
|---|---|---|---|---|---|---|---|---|---|---|
| | 112 | 114 | 115 | 116 | 117 | 118 | 119 | 120 | 122 | 124 |
| $^{114}$Sn | 0.4 | 63.2 | 0.9 | 10.6 | 3.4 | 8.4 | 2.6 | 8.8 | 0.9 | 0.9 |
| $^{120}$Sn | <0.01 | <0.01 | <0.01 | 0.04 | 0.06 | 0.10 | 0.12 | 99.6 | 0.05 | 0.02 |

**Table 2**
List of the identified residues in the proton-induced reactions of $^{114}$Sn, $^{120}$Sn targets and their spectroscopic characteristics.

| Reaction | Q value (MeV) | $E_{th}$ (MeV) | $T_{1/2}$ | $E_\gamma$ (keV), | $I_\gamma$(%) |
|---|---|---|---|---|---|
| $^{114}$Sn(p,α)$^{111g}$In | 2.69 | 0 | 2.805 d | 171.28 | 90.7 |
| | | | | 245.39 | 94.1 |
| $^{114}$Sn(p,α)$^{111m}$In | 2.69 | 0 | 7.7 min | 537.22 | 87 |
| $^{114}$Sn(p,pn)$^{113}$Sn | -10.31 | 10.39 | 115.1 d | 391.71 | 64.97 |
| $^{114}$Sn(p,d)$^{113}$Sn | -8.08 | 8.15 | | | |
| $^{114}$Sn(p,2n)$^{113}$Sb | -14.99 | 15.12 | 6.67 min | 332.4 | 14.8 |
| $^{120}$Sn(p,n)$^{120m}$Sb | -3.64 | 3.49 | 5.76 d | 197.3 | 87 |
| | | | | 1023.1 | 99.4 |
| $^{120}$Sn(p,n)$^{120g}$Sb | -3.64 | 3.49 | 15.89 min | 703.8 | 0. 149 |
| | | | | 988.6 | 0.063 |
| $^{120}$Sn(p,α)$^{117m}$In | 2.71 | 0 | 116.2 min | 315.302 | 19.1 |
| $^{120}$Sn(p,α)$^{117g}$In | 2.71 | 0 | 43.2 min | 553 | 100 |

## 3. Data analysis

In Table 3 we present our measurements of reaction cross-sections $\sigma$, which were determined as

$$\sigma = \frac{A_{obs} \lambda \frac{t_{3r}}{t_{3l}}}{\Phi N_{nucl}\, \varepsilon\, I_\gamma\, (1-e^{-\lambda t_1})e^{-\lambda t_2}(1-e^{-\lambda t_{3r}})} \quad (1).$$

Here, $A_{obs}$ is the observed number of γ-rays under the photo-peak, $\lambda$ is the decay constant, $\varepsilon$ is the detector efficiency, $N_{nucl}$ is the number of target nuclei per area, $\Phi$ is the proton flux, $I_\gamma$ is the intensity of the product gamma line, $t_1$ is the irradiation time, $t_2$ is the time between the end of the bombardment and the beginning of the measurement, $t_{3r}$ is the measurement real time, and $t_{3l}$ is the measurement live time [32-34].

The actual energy of the external beam may differ from the nominal beam energy [35]. To determine the actual energy and the beam intensity, we placed a thin (20 μm) copper monitor foil in front of the stack. The ratio of the $^{63}$Cu(p,n)$^{63}$Zn and $^{63}$Cu(p,2n)$^{62}$Zn monitor reaction cross sections was calculated using eq. (2):

$$\frac{\sigma_{^{62}Zn}}{\sigma_{^{63}Zn}} = \frac{A_{^{62}Zn}\left(1-e^{-\lambda_{^{63}Zn} t_1}\right)}{A_{^{63}Zn}\left(1-e^{-\lambda_{^{62}Zn} t_1}\right)} \quad (2),$$

where $A_{^{62}Zn}$, $\lambda_{^{63}Zn}$ and $A_{^{63}Zn}$, $\lambda_{^{62}Zn}$ are activities and decay constants of $^{62}$Zn and $^{63}$Zn, respectively [36]. The primary beam energy was estimated to be (18 ± 0.2) MeV by comparing the cross section ratio obtained from eq.(2) to the recommended cross-sections from IAEA [37]. We also calculate the beam intensity from monitor reactions from eq. (1). The proton flux in each monitor target was determined using the recommended cross sections of $^{nat}$Cu(p,x)$^{62,63,65}$Zn reactions from [37].

The average proton energy in each layer of the stack (defined as the mean of incoming $E_{in}$ and outgoing $E_{out}$ energies from the layer) and their uncertainties ($\Delta E = (E_{in} - E_{out})/2$) were calculated using SRIM-2013 [38]. The total energy uncertainties for each layer also include uncertainties from the primary beam energy (0.2 MeV) and target thicknesses (5%).

When calculating reaction cross section uncertainties, we add uncertainties from the following sources in quadrature: statistical errors on counts (0.5-16%), uncertainties in the flux (up to 7%), detector efficiency (4%), and target thicknesses (5%). The resulting cross section uncertainties are in the range of 9–19%.

## 4. Results and discussion

In figures 2- 6 we show the comparison of our measured cross-sections (given in Table 3) with published experimental data [4-5, 8-14, 17, 20, 23, 25] and theoretical values obtained using TALYS1.95 code [39].

**Table 3**
Cross-sections of the radionuclides formed in the proton-induced reactions of $^{114}$Sn, $^{120}$Sn targets at different energies.

| Energy (MeV) | Cross-section (mb) | | | |
| --- | --- | --- | --- | --- |
| | $^{114}$Sn(p,α)$^{111(m+g)}$In | $^{114}$Sn(p,α)$^{111m}$In | $^{114}$Sn(p,x)$^{113(m+g)}$Sn (cum) | $^{114}$Sn(p,2n)$^{113}$Sb |
| 17.34 ± 0.89 | 7.98 ± 0.72 | 0.58 ± 0.11 | 281 ± 25 | 21.58 ± 4.10 |
| 15.79 ± 0.81 | 6.59 ± 0.59 | – | 138 ± 13 | – |
| 14.13 ± 0.72 | 3.96 ± 0.38 | – | 26.49 ± 2.91 | – |
| 12.13 ± 0.64 | 2.52 ± 0.23 | – | 2.65 ± 0.50 | – |
| 10.52 ± 0.54 | 1.30 ± 0.19 | – | – | – |
| 8.46 ± 0.50 | 0.54 ± 0.09 | – | – | – |
| Energy (MeV) | Cross-section (mb) | | | |
| | $^{120}$Sn(p,n)$^{120m}$Sb | $^{120}$Sn(p,n)$^{120g}$Sb | $^{120}$Sn(p,α)$^{117m}$In | $^{120}$Sn(p,α)$^{117g}$In |
| 16.57 ± 0.85 | 23.34 ± 2.59 | 108 ± 15 | 1.20 ± 0.22 | 5.22 ± 0.46 |
| 14.97 ± 0.77 | 39.66 ± 3.58 | 131 ± 18 | 1.10 ± 0.11 | 3.52 ± 0.33 |
| 13.27 ± 0.69 | 49.08 ± 4.42 | 215 ± 20 | 0.76 ± 0.07 | 1.82 ± 0.16 |
| 11.48 ± 0.61 | 71.56 ± 10.73 | 551 ± 55 | 0.50 ± 0.05 | 1.08 ± 0.11 |
| 9.51 ± 0.51 | 46.78 ± 7.03 | – | 0.17 ± 0.03 | 0.46 ± 0.09 |
| 7.31 ± 0.41 | 14.56 ± 2.18 | – | – | – |

## 4.1. Reactions on the $^{114}$Sn target

$^{111}$In is directly produced from the $^{114}$Sn(p,α) reaction which has a Coulomb barrier of 7.04 MeV and no threshold. The isomeric state of $^{111}$In ($I^p$=1/2$^-$, $T_{1/2}$ = 7.7 min) transits (IT = 100%) to the ground state ($I^p$=9/2$^+$, $T_{1/2}$ = 2.805 d), which further decays (ε: 100%) to stable $^{111}$Cd. Because of its short half-life we only managed to estimate the cross-section of the isomeric state $^{111m}$In for the target γ-spectra of which were measured first.

$^{114}$Sn(p,α)$^{111(m+g)}$In reaction cross-sections were measured on a 63.2% enriched $^{114}$Sn target using interference-free γ-lines $E_\gamma$ = 171.28 keV ($I_\gamma$ = 90.7%) and $E_\gamma$ = 245.39 keV ($I_\gamma$ = 94.1%). The only other reaction that contributes to $^{111}$In production from our is $^{112}$Sn(p,2p)$^{111}$In ($E_{th}$ = 7.62 MeV). Since $^{112}$Sn content in the target was only 0.4%, and $^{112}$Sn(p,2p)$^{111(m+g)}$In reaction cross section is small (≤0.5 mb in discussed energies region according to TALYS1.95) the contribution of this reaction can be neglected. Reactions on other tin isotopes are not considered due to their high thresholds. Fig. 2 shows the measured cross-section of the discussed reaction along with Talys1.95 simulation results. Talys1.95 data are in a good agreement with the measured data. We didn't find any experimental results for the comparison.

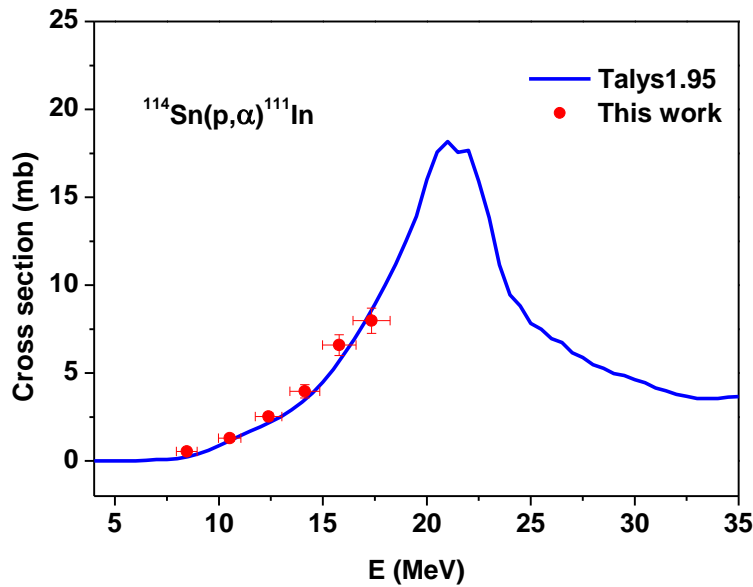

**Fig. 2.** Excitation function of the $^{114}$Sn(p,α)$^{111}$In reaction.

$^{113}$Sb is produced in $^{114}$Sn(p,2n) reactions with 15.12 MeV threshold energy and decays ($T_{1/2}$ = 6.67 min) to $^{113}$Sn by electron capture ($\varepsilon$ 100%). $^{114}$Sn(p,2n)$^{113}$Sb reaction cross-section was measured using the interference-free γ-line $E_\gamma$ = 332.41 keV ($I_\gamma$ = 14.8%). The contribution of the side reaction $^{112}$Sn(p,γ)$^{113}$Sb does not exceed the uncertainty of measured cross section for the reaction $^{114}$Sn(p,2n)$^{113}$Sb ($^{112}$Sn content in the target is 0.4%, and TALYS cross-section of $^{112}$Sn(p,γ)$^{113}$Sb reaction at 17 MeV is 2.12 mb). Fig.3 shows $^{114}$Sn(p,2n)$^{113}$Sb reaction cross sections. For this reaction, there is one additional measurement available [26]; neither of the experimental points is in agreement with the calculations, and additional measurements are required to construct the full shape of the excitation function.

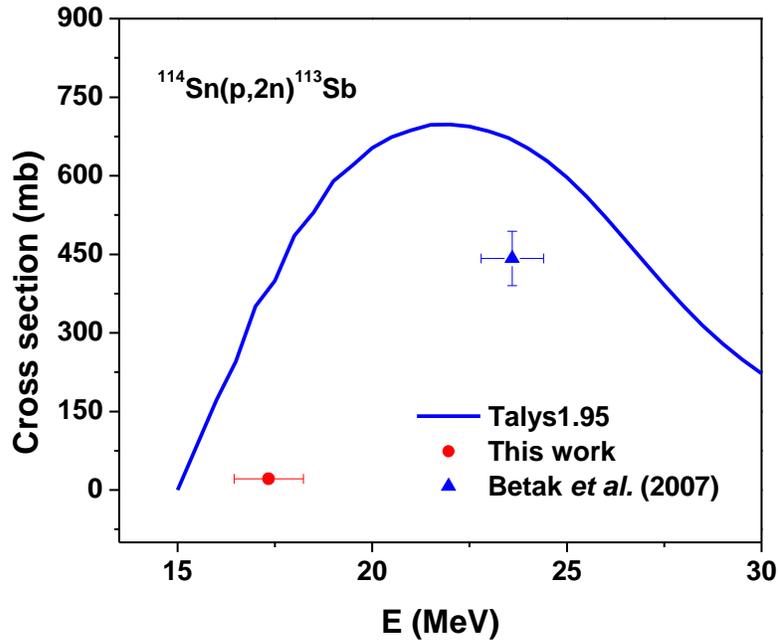

**Fig.3**. Excitation function of the $^{114}$Sn(p,2n)$^{113}$Sb reaction.

Ground state of $^{113g}$Sn ($I^p=1/2^+$, $T_{1/2}$ = 115.09 d) forms through direct production with (p,pn) reaction, decay (IT 91%) of the metastable $^{113m}$Sn state ($I^p = 7/2^+$, $T_{1/2}$ = 21.4 min)), and decay of the parent $^{113}$Sb. $^{113}$Sn formation chain is presented in Fig 3.

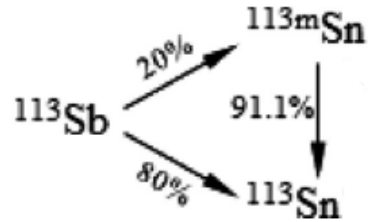

**Fig. 4.** $^{113}$Sn formation chain.

Full cumulative cross-sections were measured after the total decay of the two short-lived parents through the 391.7 keV γ-ray line ($I_\gamma$ = 64.97%) of $^{113g}$Sn. Contributions of other Sn isotopes contained in the target to the $^{113}$Sn production can be neglected: $^{115}$Sn(p,p2n)$^{113}$Sn reaction ($E_{th}$ = 9.45 MeV) can be neglected as $^{115}$Sn content in the target is only 0.9%, and $^{115}$Sn(p,x)$^{113}$Sn reaction cross section is small (≤3.5·10$^{-3}$ mb for the discussed energy region according to TALYS1.95). $^{112}$Sn(p,γ)$^{113}$Sb reaction contribution does not exceed the uncertainty of our measurement for the $^{114}$Sn(p,2n)$^{113(m+g)}$Sn(cum) reaction cross section (TALYS1.95 reaction cross sections for the energies 12.13–15.79 MeV are ≤7.01 mb). Fig. 5 shows the comparison of $^{114}$Sn(p,x)$^{113(m+g)}$Sn(cum) reaction cross-section measurements with the values from Talys1.95 simulations showing good agreement at low energies.

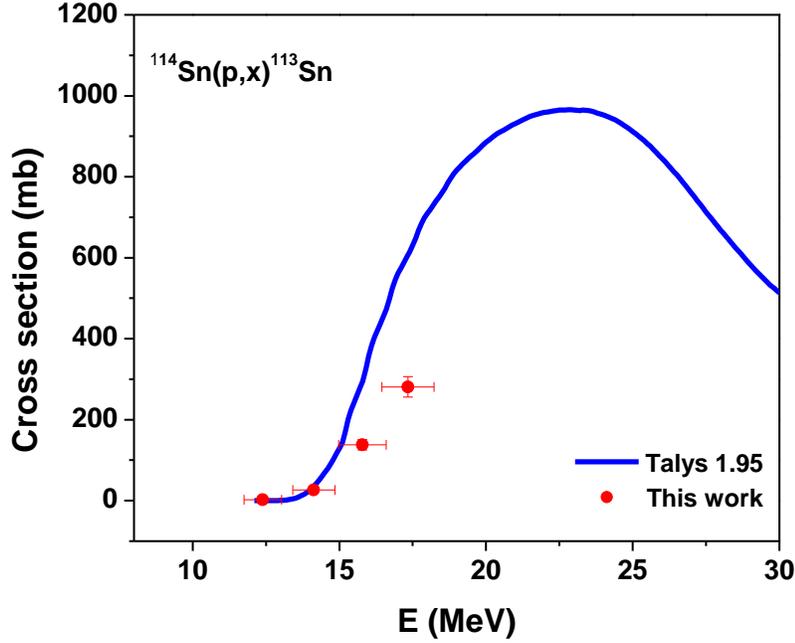

**Fig. 5.** Excitation function of the $^{114}$Sn(p,x)$^{113}$Sn.

*4.2. Reactions on the $^{120}$Sn target*

Here we measure $^{120}$Sb and $^{117}$In isotope production cross-sections. The purity of the $^{120}$Sn target used in the experiments is 99.6%, making the contributions of other tin isotopes to product yields negligibly small: $^{119}$Sn and $^{118}$Sn contents in the target are 0.12% and 0.10% and $^{119}$Sn(p,γ)$^{120}$Sb and $^{118}$Sn(p,2p)$^{117}$In reaction cross sections in the discussed energy region are ≤0.99 mb and ≤10$^{-7}$ mb, respectively.

$^{120}$Sb is produced in the $^{120}$Sn(p,n) reaction in both ground ($I^p$=8$^-$, $T_{1/2}$ = 5.76 d) and isomeric ($I^p$=1$^+$, $T_{1/2}$ = 15.89 min) states. Both states of $^{120}$Sb decay to the stable $^{120}$Sn. We measure the cross-sections of both states (summarized in Table 3) using interference-free γ-lines $E_\gamma$ = 197.3 keV ($I_\gamma$ = 87%) and $E_\gamma$ = 1023.1 keV ($I_\gamma$ = 99.4%) for $^{120}$Sn(p,n)$^{120m}$Sb reaction and $E_\gamma$ = 703.8 keV ($I_\gamma$ = 0.149%) and $E_\gamma$ = 988.6 keV ($I_\gamma$ = 0.063%) for $^{120}$Sn(p,n)$^{120g}$Sb reaction.

Fig. 6 shows the excitation functions of these reactions along with the data from Refs. [4-5, 8, 11-13] and Talys1.95 simulation results. Note that Refs. [4-5] used a $^{nat}$Sn target so for a comparison to the rest of the measurements we adjusted their results to a pure $^{120}$Sn target; these measurements may also include a small contribution from $^{119}$Sn(p,γ)$^{120m}$Sb cross section.

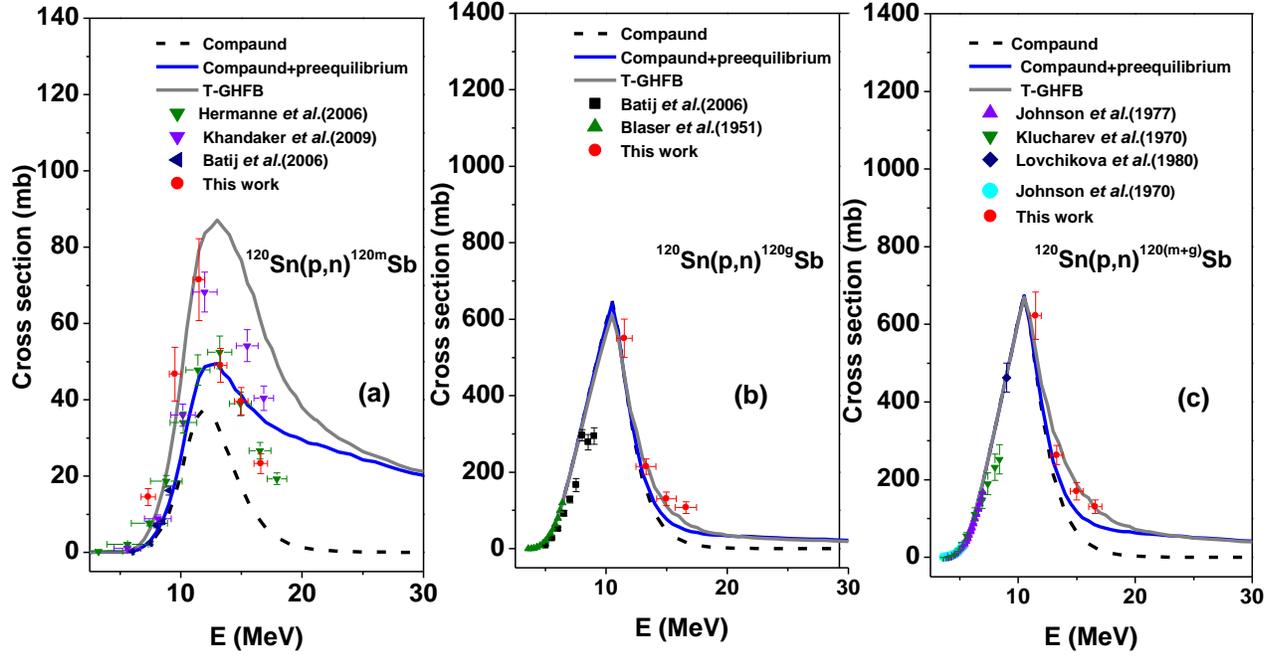

**Fig.6.** Excitation function of the (a) $^{120}$Sn(p,n)$^{120m}$Sb, (b) $^{120}$Sn(p,n)$^{120g}$Sb, (c) $^{20}$Sn(p,n)$^{120(m+g)}$Sb reactions.

All data for $^{120}$Sn(p,n)$^{120(m+g)}$Sb and $^{120}$Sn(p,n)$^{120g}$Sb reactions agree well with each other and with Talys1.95 calculations. Some discrepancies are present for the $^{120}$Sn(p,n)$^{120m}$Sb reaction at energies higher than the peak energy.

According to TALYS prediction, compound nucleus reaction plays a crucial role at low incident energies. Besides this, pre-equilibrium processes have a sizable contribution to reaction cross sections for incident energies between 10 and (at least) 200 MeV [31]. For the $^{120}$Sn(p,n)$^{120m,g}$Sb reaction we performed TALYS1.95 calculations with two models, pure compound model and compound + pre-equilibrium model. Pre-equilibrium processes play a role in formation of both $^{120m}$Sb and $^{120g}$Sb, but are much more pronounced for the metastable state as seen in Fig.5 at high energies.

$^{120}$Sn(p,α) reaction results in both the ground and isometric states of $^{117}$In. The isomeric state ($I^p = 1/2^-$, $T_{1/2} = 116.2$ min), transits to the ground state ($I^p=9/2^+$, $T_{1/2} = 43.2$ min) with IT =

47.1%. The independent cross-section of the isomeric state ($^{117m}$In) was determined using the line $E_\gamma = 315.302$ keV ($I_\gamma = 19\%$).

We calculate $^{117g}$In cross-section as a sum of direct production from $^{120}$Sn(p,α) reaction and through decay of $^{117m}$In during and after the irradiation.

The independent cross sections of daughter nuclei $^{117g}$In was calculated by the following relation [42].

$$\sigma_d = \frac{\lambda_d}{(1 - e^{-\lambda_d t_1})e^{-\lambda_d t_2}(1 - e^{-\lambda_d t_3})} \times$$

$$\left[ \frac{A_{obs}}{\Phi N_{nucl}\, \varepsilon\, I_\gamma} - \sigma_p f \frac{\lambda_p \lambda_d}{\lambda_d - \lambda_p} \left( \frac{(1-e^{-\lambda_p t_1})e^{-\lambda_p t_2}(1-e^{-\lambda_p t_3})}{\lambda_p^2} - \frac{(1-e^{-\lambda_d t_1})e^{-\lambda_d t_2}(1-e^{-\lambda_d t_3})}{\lambda_d^2} \right) \right] \quad (3),$$

where $\lambda_p$ and $\lambda_d$ are the parent and daughter decay constants, $f$ specifies the fraction of parent nuclei decaying to a daughter nucleus, and the rest of the symbols are the same as in eq.(1).

Fig. 7 shows the excitation functions of $^{120}$Sn(p,α)$^{117m}$In, $^{120}$Sn(p,α)$^{117g}$In, $^{120}$Sn(p,α)$^{117(m+g)}$In reactions. We found only one publication with the experimental results for $^{120}$Sn(p,α)$^{117m}$In reaction [18] which agree with our data. All experimental data are shifted towards lower energies compared to TALYS1.95 calculation.

In TALYS1.95, the cross section of the (p,α) reaction is given by the sum of compound and pre-equilibrium components. The latter is described by the exciton model. Pick-up and knock-out mechanisms, which play an important role for (p,α) reactions [2], are not covered by the exciton model. A specially developed phenomenological model is included in TALYS1.95 for covering these processes [39].

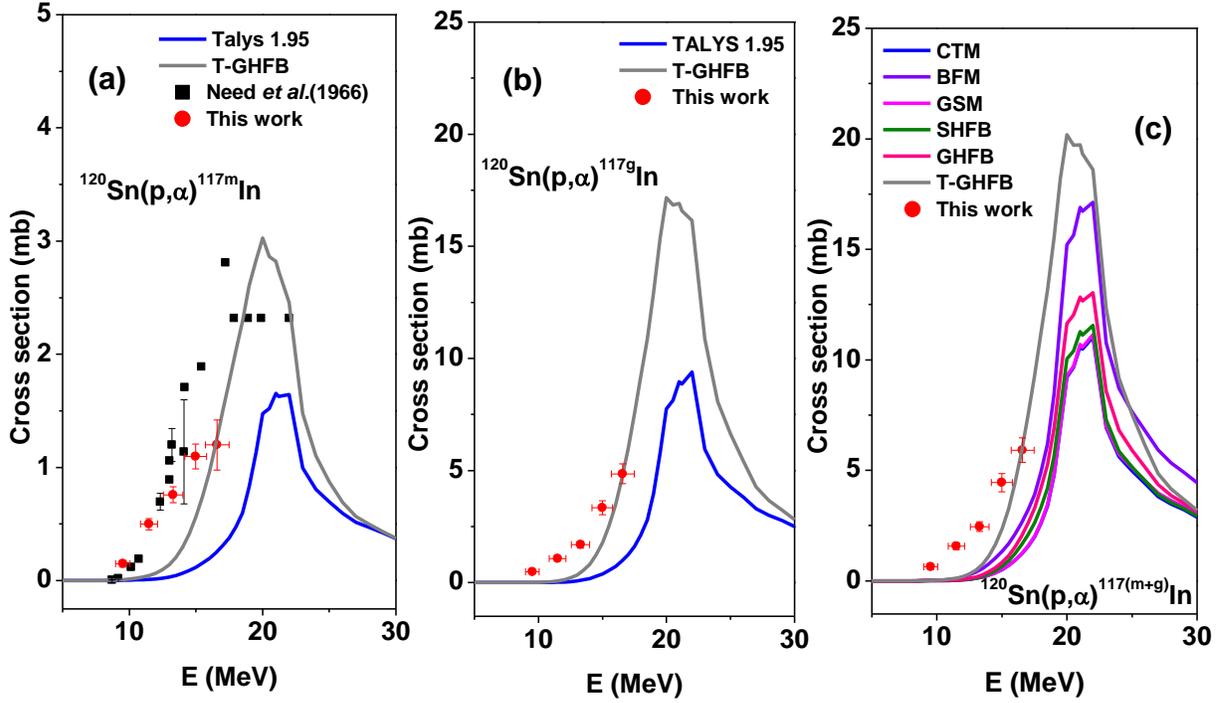

**Fig. 7.** Excitation functions of the (a) $^{120}$Sn(p,α)$^{117m}$In, (b) $^{120}$Sn(p,α)$^{117g}$In, (c) $^{120}$Sn(p,α)$^{117(m+g)}$In reactions.

In this paper, we have considered how the excitation function of the reaction (p,α) changes depending on the nuclear level model. Nuclear level densities are the most important values for compound nucleus decay width calculations in the statistical model. TALYS1.95 includes six different models for nuclear levels, and we made calculations with all of them. The six models are Constant Temperature + Fermi gas model (CTM) (this model is default in TALYS), Back-shifted Fermi gas Model (BFM), Generalised Superfluid Model (GSM), Skyrme-Hartree-Fock-Bogoluybov level densities from numerical tables (SHFB), Gogny-Hartree-Fock-Bogoluybov level densities from numerical tables (GHFB), and Temperature-dependent Gogny-Hartree-Fock-Bogoluybov level densities from numerical tables (T-GHFB) [39]. The results of calculations for various models differ significantly from each other, with T-GHFB results for the $^{120}$Sn(p,α) reaction being closest to experimental ones (Fig. 7). The same model is consistent with the experiments for the $^{120}$Sn(p,n)$^{120g}$Sb reaction but overestimates the cross section of metastable $^{120m}$Sb (Fig. 6). For reactions on the $^{114}$Sn target we performed calculations only with the CTM model (Fig. 2,3,5) since the agreement with the experiment was good.

## 5. Isomeric cross section ratio

Measurements of isomeric cross section ratios (IR) are important for studies of nuclear structure, especially the level density and the discrete level structure of residual nuclei [9, 43]. It has been shown that IRs depend on several factors, such as the type and energy of the projectile, type of the emitted particle, spin of the target nucleus, and spins of the ground and isomeric states [40, 44]. Isomeric cross-section ratios measurement is important for properly choosing the input model parameters [45, 46].

We measure cross-sections of two isomeric pairs $^{120m,g}$Sb and $^{117m,g}$In formed in $^{120}$Sn(p,n) and $^{120}$Sn(p,α) reactions. IR is defined as the cross-section ratio of the higher spin state to the lower one. Measured and calculated (via TALYS1.95) IRs of the $^{120m,g}$Sb pair are presented in Fig. 8. TALYS1.95 data are consistent with low energy data of Ref. [8] and also with our data, with the exception of the highest energy measurement.

IR dependence on the incident proton energy for the $^{117m,g}$Sb pair is presented in Fig.9. Our data is in good agreement with TALYS1.95 calculations while the data of [18] are inconsistent with both.

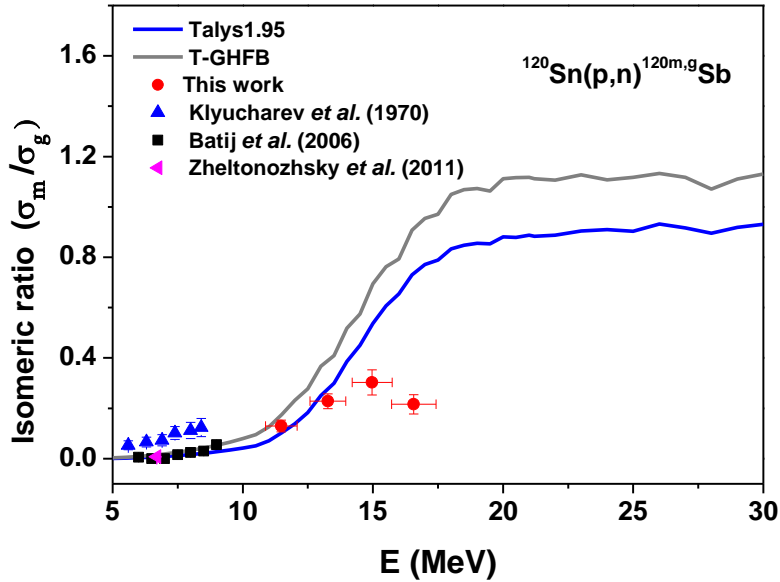

**Fig. 8.** Isomeric ratio ($\sigma_m/\sigma_g$) dependence on the incident protons energies for the $^{120m,g}$Sb.

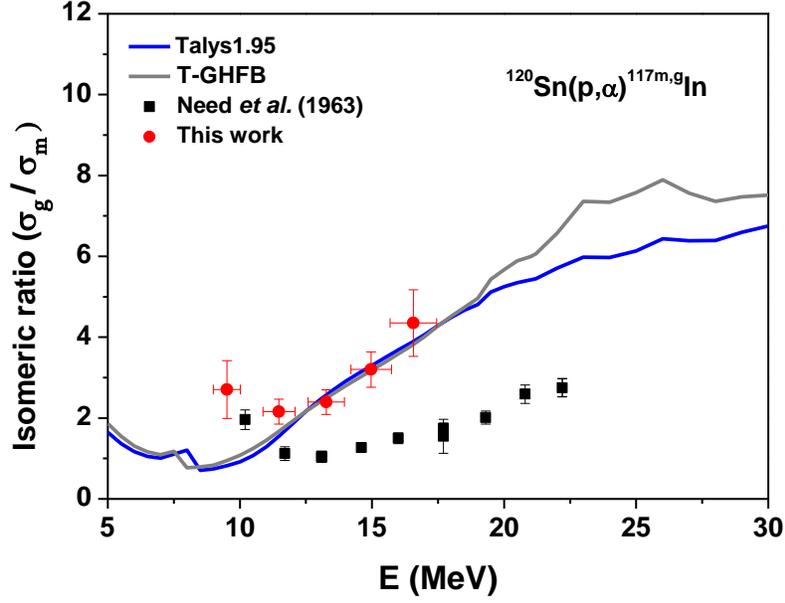

**Fig. 9.** Isomeric ratio $\sigma_g/\sigma_m$ dependence on the incident protons energies for the $^{117m,g}$In.

## 6. Conclusions

We measured $^{114}$Sn(p,α)$^{111(m+g)}$In, $^{114}$Sn(p,pn)$^{113(m+g)}$Sn, $^{114}$Sn(p,2n)$^{113}$Sb, $^{120}$Sn(p,n)$^{120m,g}$Sb, $^{120}$Sn(p,α)$^{117m,g}$In reaction cross sections and compared them to published experimental data and TALYS1.95 numerical calculations to check the predictive ability of TALYS1.95 for tin isotopes.

TALYS1.95 provides a good description for most discussed reactions. The largest discrepancy between the experimental and calculated data was noted for the $^{120}$Sn(p,α)$^{117}$In reaction. For the $^{120}$Sn target calculations with all available level density models included in TALYS1.95 were performed and it was shown that the results strongly depend on the choice of the model for the $^{120}$Sn(p,α)$^{117}$In and $^{120}$Sn(p,n)$^{120m}$Sb reactions.

For the $^{120}$Sn(p,α)$^{117}$In reaction temperature-dependent Gogny-Hartree-Fock-Bogoluybov level density gives the best agreement to the experimental values, but some discrepancy between the calculations and experiment still remains.

For the $^{120}$Sn(p,n) reaction the choice of the nuclear level density model is essential for the metastable state ($^{120m}$Sb) production and much smaller for the ground state ($^{120g}$Sb). However, due to the scatter of the available experimental data, it is impossible to unambiguously conclude that either of the two discussed models has the best predictive ability.

Isomeric cross section ratios of $^{120m,g}$Sb and $^{117m,g}$In pairs formed in $^{120}$Sn(p,n) and $^{120}$Sn(p,α) reactions are mostly well captured by Talys1.95 calculations in the discussed energy region.

**Acknowledgements**